\documentclass[12pt,preprint]{aastex}
\usepackage{psfig}
\def\e{\epsilon}
\def\I{I_{\rm d}}
\def\rs{\rho_{\rm s}}

\begin{document}

\title{Radiative Transfer in Obliquely Illuminated Accretion Disks}

\author
{Dimitrios Psaltis}
\affil{School of Natural Sciences, Institute for Advanced Study,
Einstein Dr.,\\ Princeton, NJ 08540}
\email{dpsaltis@ias.edu}

\begin{abstract}
 The illumination of an accretion disk around a black hole or neutron
 star by the central compact object or the disk itself often
 determines its spectrum, stability, and dynamics.  The transport of
 radiation within the disk is in general a multi-dimensional,
 non-axisymmetric problem, which is challenging to solve.  Here, I
 present a method of decomposing the radiative transfer equation that
 describes absorption, emission, and Compton scattering in an
 obliquely illuminated disk into a set of four one-dimensional
 transfer equations.  I show that the exact calculation of the
 ionization balance and radiation heating of the accretion disk
 requires the solution of only one of the one-dimensional equations,
 which can be solved using existing numerical methods.  I present a
 variant of the Feautrier method for solving the full set of
 equations, which accounts for the fact that the scattering kernels in
 the individual transfer equations are not forward-backward symmetric.
 I then apply this method in calculating the albedo of a cold,
 geometrically thin accretion disk.
\end{abstract}


\section{INTRODUCTION}

The structure and dynamics of an accretion disk around a neutron star
or a black hole is often strongly influenced by its irradiation from
the compact object or the disk itself. In galactic sources, for
example, the optical emission (see, e.g., de Jong, van Paradijs, \&
Augusteijn 1996), the stability (see, e.g., van Paradijs 1996; King,
Kolb, \& Burderi 1996), and the warping of the accretion disk (Pringle
1996; Wijers \& Pringle 1999) are probably determined by X-ray
irradiation from the central source. In extragalactic sources, the
observed broad iron emission lines (see Nandra 1997 for a review), the
weak or absent Lyman edges (Sincell \& Krolik 1997; Antonucci 1999),
as well as the disk warping (Pringle 1996) may also be caused by X-ray
irradiation of the accretion disk.

Calculating the structure even of a non-illuminated accretion disk is
not trivial, because of the complicated physical processes involved,
such as the non-LTE character of the transport of radiation (see,
e.g., Hubeny \& Hubeny 1998), and assumptions regarding the vertical
profile of the viscous heating (see, e.g., Laor \& Netzer 1989). The
problem becomes even more difficult when illumination is taken into
account, both because additional physical process, such as
photoionization (see, e.g., Raymond 1993, Ko \& Kallman 1994), become
dominant, and because of the multidimensional character of the
problem.

The transport of radiation through an accretion disk illuminated by a
distant source is in general two-dimensional and not axisymmetric. The
spectrum of the reflected radiation, when only Compton scattering is
taken into account, has often been solved in terms of the distribution
of photon escape probabilities from the reflecting medium (e.g.,
Lightman \& Rybicki 1980) or using Green's functions (e.g., Poutanen,
Nagendra, \& Svenson 1996). When absorption and emission processes are
taken into account, the problem is often solved assuming
plane-parallel symmetry (e.g., Sincell \& Krolik 1997), or employing
variants of the $\Lambda$-iteration (e.g., Basko, Sunyaev, \&
Titarchuk 1974; Matt, Fabian, \& Ross 1993) and Monte-Carlo methods
(e.g., George \& Fabian 1991).

Solving directly the radiative transfer equation in two spatial
dimensions is challenging, especially when the energies of the photons
change by Compton scattering and the problem becomes non-local both in
space and in photon energy (see, e.g., Mihalas, Auer, \& Mihalas 1978; Auer
\& Paletou 1994; Dykema, Klein, \& Castor 1996; Dullemond \& Turolla 2000;
Busche \& Hillier 2000 for methods of solving problems in two spatial
dimensions). In general, a problem that is two-dimensional in
coordinate space requires the solution of the radiative transfer
equation in five dimensions, two in coordinate space and three in
photon momentum space. In the case of illumination of a geometrically
thin accretion disk, however, the problem can be simplified
significantly. Because the photon mean free paths in the radial and
azimuthal directions are much smaller than any characteristic length
scale in the accretion flow, a geometrically thin accretion disk can
be decomposed into a finite number of plane-parallel, obliquely
illuminated slabs. Solving the radiative transfer equation for each
slab requires only four dimensions, one in coordinate space and three
in photon momentum space. When the angular dependence of the
interaction cross section between photons and matter can be expanded
into a finite number of Legendre polynomials, the problem can be
simplified even further (Chandrasekhar 1960). In this case, the
radiative transfer equation is equivalent to a finite number of
equations over one dimension in coordinate space and two dimensions in
photon momentum space, i.e., to a finite number of radiative transfer
problems in one spatial dimension. These equations can be solved in
general using existing numerical methods.

In this paper I derive the system of equations (reduced to one
spatial dimension) that describe, in an obliquely illuminated slab,
isotropic absorption and emission, as well as Compton scattering to
first order in $\epsilon/m_{\rm e}$ and $T_{\rm e}/m_{\rm e}$, where
$\epsilon$ is the photon energy and $T_{\rm e}$ and $m_{\rm e}$ are
the electron temperature and rest mass. I then explore a variant of
the Feautrier method for solving the resulting transfer equations, in
which the scattering kernels do not have forward-backward symmetry
(see Milkey, Shine, \& Mihalas 1975).  I illustrate the use of the
derived equations and method of solution by solving simple problems
related to the albedo of a cold disk. More detailed calculations of
the coupled radiation and gas properties using the approach described in
this paper will be reported elsewhere.

\section{RADIATIVE TRANSFER}

I solve the radiative transfer problem in a plane-parallel slab
illuminated by some arbitrary, external source of photons. I describe
all physical quantities in this system using an orthogonal, Cartesian
reference frame, with its $z$-axis parallel to the finite dimension of
the slab. I also set $c=k_{\rm B}=h=1$, where $c$ is the speed of
light, $k_{\rm B}$ is Boltzman's constant, and $h$ is Planck's
constant.

I assume that the electrons in the slab have density $n_{\rm e}(z)$
and temperature $T_{\rm e}(z)\ll m_e$, where $m_e$ is the electron
rest mass. I also assume that the illuminating radiation is a parallel
pencil beam of net flux $\pi F$. Neglecting induced processes, the
radiative transfer problem is linear and therefore the effect on any
arbitrary illumination pattern can be computed by summing the
solutions obtained for each plane-parallel pencil beam.  The direction
of illumination is described by the vector $\vec{r}=\sin\theta_{\rm
i}\sin\phi_{\rm i} \hat{x}+ \sin\theta_{\rm i}\cos\phi_{\rm i}
\hat{y} + \cos\theta_{\rm i} \hat{z}$, where
$\theta_{\rm i}$ and $\phi_{\rm i}$ are the directional angles.

I describe the radiation field in terms of the monochromatic specific
intensity $I(z,\mu,\phi,\epsilon)$, where $z$ is the distance from the
edge of the slab, $\mu\equiv\cos\theta$ and $\phi$ are the directional
angles of the propagation vector $\hat{l}$, and $\epsilon$ is the
photon energy.  Because I study non-polarized radiation, I have
suppressed the dependence of the specific intensity on polarization
mode. I assume that absorption and emission in the slab are isotropic
and denote the absorption coefficient by $\chi(z,\e)$ and the source
function by $S(z,\e)$.

Keeping only terms to first order in $\e/m_e$ and $T_e/m_e$, the
radiative transfer equation that describes absorption, emission, and
Compton scattering can be written as (Pomraning 1973)
 \begin{eqnarray}
 \hat{l}\cdot \nabla I(z,\hat{l},\e)&=&
 \chi(z,\e) [S(z,\e)-I(z,\hat{l},\e)] 
 - n_e \sigma_{\rm T}\left(1-2\frac{\e}{m_e}\right) 
 I(z,\hat{l},\e)\nonumber\\
& &\qquad\qquad+n_e\sigma_{\rm T} \int d\Omega'
 \sum_{n=0}^3 \left(\frac{2n+1}{4\pi}\right)
 P_n(\hat{l}\cdot\hat{l}')S_n I(z,\hat{l}',\e)\;,
 \label{genRTE}
 \end{eqnarray}
 where $\sigma_{\rm T}$ is the angle-integrated cross
section for Thomson scattering, $d\Omega'$ is the
solid-angle element around $\hat{l}'$,
$P_n(\hat{l}\cdot \hat{l}')$ is the Legendre
polynomial of order $n$, and
 \begin{eqnarray}
 S_0 & = & 1-\frac{\e}{m_e}\left(1-
    \e\frac{\partial}{\partial \e}\right)-
 \frac{T_e}{m_e}\left(2\e\frac{\partial}{\partial \e}
 -\e^2\frac{\partial^2}{\partial \e^2}\right)
  \;,\nonumber\\
 S_1 & = & \frac{2}{5}\left[\frac{\e}{m_e}\left(1-
    \e\frac{\partial}{\partial \e}\right)-
 \frac{T_e}{m_e}\left(1-2\e\frac{\partial}{\partial \e}
 +\e^2\frac{\partial^2}{\partial \e^2}\right)\right]
  \;,\nonumber\\
 S_2 & = & \frac{1}{10} \left[1-\frac{\e}{m_e}\left(1-
    \e\frac{\partial}{\partial \e}\right)-
 \frac{T_e}{m_e}\left(6+2\e\frac{\partial}{\partial \e}
 -\e^2\frac{\partial^2}{\partial \e^2}\right)\right]
  \;,\nonumber\\
 S_3 & = & \frac{3}{70}\left[\frac{\e}{m_e}\left(1-
    \e\frac{\partial}{\partial \e}\right)+
 \frac{T_e}{m_e}\left(4+2\e\frac{\partial}{\partial \e}
 -\e^2\frac{\partial^2}{\partial \e^2}\right)\right]\;.
 \label{Sn}
 \end{eqnarray}
 Defining the energy-dependent optical depth as
 \begin{equation}
 d\tau(\epsilon) \equiv
-\left[\left(1-\frac{2\epsilon}{m_e}\right) 
   n_e \sigma_{\rm T}+\chi \right]dz\;,
 \end{equation}
 the relative importance of absorption and scattering
as
 \begin{equation}
 \rho_{\rm a}=\frac{\chi}{(1-2\epsilon/m_e) 
     n_e \sigma_{\rm T}+\chi}
 \end{equation}
 and
 \begin{equation}
 \rho_{\rm s}=\frac{n_e \sigma_{\rm T}}
   {(1-2\epsilon/m_e) n_e \sigma_{\rm T}+\chi}\;,
 \end{equation}
 and the redistribution function in the scattering
Kernel as 
 \begin{equation}
 p(\hat{l}\cdot\hat{l}')\equiv \sum_{n=0}^3
   \varpi_n P_n(\hat{l}\cdot\hat{l}')
 \label{redis}
 \end{equation}
 with $\varpi_n=(2n+1)\rho_{\rm s}S_n$, the radiative
transfer equation~(\ref{genRTE}) becomes
 \begin{eqnarray}
 \mu\frac{\partial}{\partial \tau} I(\tau,\mu,\phi,\e)
 &=& I(\tau,\mu,\phi,\e)-\rho_{\rm a}S(\tau,\e)
 -\frac{1}{4\pi} \int d\Omega'
     p(\hat{l}\cdot\hat{l}')I(\tau,\mu',\phi',\e)\;.
 \label{4dRTE}
 \end{eqnarray}
Equation~(\ref{4dRTE}) is a second-order, intergrodifferential
equation in a four-dimensional phase space and is equivalent to a
system of four equations in a three-dimensional phase space (see
Chandrasekhar 1960, \S 48.1). This is true because the expansion in
Legendre polynomials of the redistribution function~(\ref{redis})
terminates after the first four terms.

Following Chandrasekhar (1960), I will write the radiative transfer
equation~(\ref{4dRTE}) in terms of the specific intensity
$\I(\tau,\mu,\phi,\e)$ of the diffuse radiation field, i.e., of the
photons that have interacted with the gas at least once. Expanding the
specific intensity of the diffuse radiation field as
 \begin{equation}
 \I(\tau,\mu,\phi,\e)=\sum_{m=0}^3 \I^m(\tau,\mu,\e)
   \cos\left[m(\phi_{\rm i}-\phi)\right]\;,
 \label{expans}
 \end{equation}
where $\phi_{\rm i}$ defines the plane of
illumination, the transfer equation for the diffuse
field becomes equivalent to the system of equations
(cf.\ Chandasekhar 1960, \S 48.1)
 \begin{eqnarray}
 \mu\frac{\partial}{\partial \tau} \I^m(\tau,\mu,\e)
 & = & \I^m(\tau,\mu,\e) 
 - \delta_0^m \rho_{\rm a}S(\tau,\e)
 -\frac{1}{2} \sum_{n=m}^3 \varpi_n^m P_n^m (\mu)
  \int_{-1}^1 P_n^m (\mu')\I^m(\tau,\mu',\e)d\mu'
 \nonumber\\
 & & \qquad\qquad
 -\left[\frac{1}{4}(2-\delta_0^m) \sum_{n=m}^3
  \varpi_n^m (-1)^{m+n} P_n^m(\mu) P_n^m(\mu_{\rm i})\right]
 Fe^{-\tau/\mu_{\rm i}}\;,
 \label{gensys}
 \end{eqnarray}
 where $\delta_0^m$ is Kronecker's delta, 
 \begin{equation}
 \varpi_n^m\equiv \varpi_n \frac{(n-m)!}{(n+m)!}\;,
 \end{equation}
 and $P_n^m(\mu)$ are the associated Legendre's 
functions of the first kind defined by
 \begin{equation}
 P_n^m(\mu)=(-1)^m(1-\mu^2)^{m/2} \frac{d^mP_n(\mu)}
 {d\mu^m}\;. 
 \end{equation}

Written in an explicit form, the zeroth-order equation is
 \begin{eqnarray}
 \mu\frac{\partial \I^0(\mu)}{\partial\tau}
& = & \I^0(\mu)
 -\left[\frac{1}{2}S_0 - \frac{5}{8}S_2(3\mu^2-1)
     \right]\rs\int_{-1}^1 \I^0(\mu')d\mu'\nonumber\\
& &\qquad
 -\left[\frac{3}{2}S_1\mu-\frac{21}{8}S_3 (5\mu^3-3\mu)
   \right]\rs\int_{-1}^1\mu'\I^0(\mu')d\mu'\nonumber\\
& & \qquad
 -\frac{15}{8}S_2 (3\mu^2-1)\rs\int_{-1}^1 \mu'^2
   \I^0(\mu')d\mu'
 -\frac{35}{8}S_3(5\mu^3-3\mu)\rs\int_{-1}^1
    \mu'^3 \I^0(\mu')d\mu'\nonumber\\
& & \qquad
 -\frac{1}{4} \left[
 S_0-3S_1\mu\mu_{\rm i} 
  +\frac{5}{4}S_2(3\mu^2-1)(3\mu_{\rm i}^2-1)
 -\frac{7}{4}S_3(5\mu^3-3\mu)(5\mu_{\rm 
 i}^3-3\mu_{\rm i})\right]\rs
 Fe^{\tau/\mu_{\rm i}}\;.
 \label{PDE0}
 \end{eqnarray}
 The first-order equation is
 \begin{eqnarray}
 \frac{\mu}{1-\mu^2}\frac{\partial 
{\cal L}_{\rm d}^1(\mu)}{\partial \tau} 
& = & \frac{{\cal 
       L}_{\rm d}^1(\mu)}{1-\mu^2}
    -\frac{3}{4}\left[S_1-\frac{21}{8}S_3 (5\mu^2-3)\right]
    \rs\int_{-1}^1 {\cal L}_{\rm d}^1(\mu')d\mu'\nonumber\\
& & \qquad
   -\frac{15}{4}S_2 \mu \rs\int_{-1}^1\mu'
       {\cal L}_{\rm d}^1(\mu')d\mu'
   -\frac{105}{32}S_3(5\mu^2-3)\rs\int_{-1}^1 \mu'^2
       {\cal L}_{\rm d}^1(\mu')d\mu'\nonumber\\
& & \qquad
   -\frac{3}{4}\left[S_1(1-\mu_{\rm i})^{1/2}-
 5S_2\mu\mu_{\rm i}(1
    -\mu_{\rm i}^2)^{1/2}\right.\nonumber\\
& & \qquad\qquad\qquad \left.
      +\frac{7}{8}S_3(5\mu^2-3)(5\mu_{\rm i}^2-3)
 (1-\mu_{\rm i}^2)^{1/2}\right]\rs
  Fe^{-\tau/\mu_{\rm i}}\;,
 \label{PDE1}
 \end{eqnarray}
 where ${\cal L}^1_{\rm d}(\mu)\equiv (1-\mu^2)^{1/2}I_{\rm d}^1(\mu)$. 
The second-order equation is
 \begin{eqnarray}
 \mu\frac{\partial \I^2(\mu)}{\partial \tau} 
& = & \I^2(\mu)
   -\frac{15}{16}S_2(1-\mu^2)\rs\int_{-1}^1
   \I^2(\mu')d\mu'
   \nonumber\\
& & \qquad
   -\frac{105}{16}S_3 \mu(1-\mu^2)\rs\int_{-1}^1
      \mu'\I^2(\mu')d\mu'
   +\frac{15}{16}S_2(1-\mu^2)\rs\int_{-1}^1
      \mu'^2\I^2(\mu')d\mu'\nonumber\\
& & \qquad
   +\frac{105}{16}S_3\mu(1-\mu^2)\rs\int_{-1}^1
   \mu'^3 \I^2(\mu')d\mu'
   \nonumber\\
& & \qquad
   -\frac{15}{16}\left[S_2(1-\mu^2)(1-\mu_{\rm i}^2)
   -7S_3\mu\mu_{\rm
   i}(1-\mu^2)(1-\mu_{\rm i}^2)\right]\rs
      Fe^{-\tau/\mu_{\rm i}}\;.
 \label{PDE2}
 \end{eqnarray}
 Finally, the third-order equation is
 \begin{eqnarray}
 \mu\frac{\partial \I^3(\mu)}{\partial \tau}
& = & \I^3(\mu)
   - \frac{35}{32}S_3 (1-\mu^2)^{3/2}
     \rs\int_{-1}^1 (1-\mu'^2)^{3/2}\I^3(\mu')d\mu'
   \nonumber\\
& &\qquad  -\frac{35}{32}S_3(1-\mu^2)^{3/2}(1-\mu_{\rm
   i}^2)^{3/2}
      \rs Fe^{-\tau/\mu_{\rm i}}\;.
 \label{PDE3}
 \end{eqnarray}

Equations~(\ref{PDE0})--(\ref{PDE3}) are four, second-order, partial
differential equations. Because they describe the evolution of the
diffuse radiation field, they can be solved with the following boundary 
conditions, 
 \begin{equation}
 \I^{\rm m}(\tau=0,\mu<0,\epsilon)=\I^{\rm m}(\tau_{\rm max},\mu>0,\epsilon)=0
  \;,
  \label{tau_bound}
 \end{equation}
where $\tau_{\rm max}$ is the total vertical optical depth of the slab,
and
 \begin{equation}
 \I^{\rm m}(\tau,\mu,\epsilon=0)=\I^{\rm m}(\tau,\mu,\epsilon\rightarrow\infty)
   = 0\;.
  \label{en_bound}
 \end{equation}
In practice, the vertical optical depth of an accretion disk can be very
large and, therefore, the second of boundary conditions~(\ref{tau_bound})
can be exchanged with another condition that is easier to handle numerically.

The first two moments of the specific intensity of the diffuse
radiation field, which are necessary for calculating the energy and
momentum exchange between photons and matter, can be calculated as
 \begin{equation}
 J_{\rm d}(\tau,\e)=\frac{1}{2}\int_{-1}^1
   \I^0(\tau,\e,\mu)
   d\mu
 \label{J}
 \end{equation}
 and
 \begin{equation}
 \vec{H_{\rm d}}
  (\tau,\e)=H_1(\tau,\e)\cos(\phi_{\rm i})\hat{x}
    +H_1(\tau,\e)\sin(\phi_{\rm i})\hat{y}
    +H_2(\tau,\e)\hat{z}\;,
 \label{Hvec}
 \end{equation}
 where 
 \begin{equation}
  H_1(\tau,\e)
   =\frac{1}{4}\int_{-1}^1(1-\mu^2)^{1/2}
     \I^1(\tau,\e,\mu)d\mu
 \label{H1}
 \end{equation}
 and
 \begin{equation}
 H_2(\tau,\e)=\frac{1}{4}\sum_{m=0}^3\int_{-1}^1\mu
     \I^m(\tau,\e,\mu) d\mu\;.
 \label{H2}
 \end{equation}

\section{NUMERICAL METHOD}

In \S2, I showed that the specific radiative transfer problem in two
spatial dimensions described by equation~(\ref{genRTE}) has been
reduced to four problems in one spatial dimension each, which are
easier to solve. However, solving the latter problems still requires
special care, for a number of reasons. First, the interaction of the
illuminating radiation with the disk material takes place in the
outermost layers of the slab, which are optically thin. As a result,
the method of solution of the transfer equation must be accurate in
the limit of low optical depth.  Second, in a typical
disk-illumination problem, the interaction of the illuminating
radiation with the disk material is dominated by true-absorption at
low photon energies but is scattering-dominated at high
photon-energies.  For this reason, simple $\Lambda$-iteration
procedures are not adequate and either accelerated iterative
procedures (e.g., the accelerated $\Lambda$-iteration or the method of
variable Eddington factors) or other non-iterative procedures (e.g.,
the Feautrier method) must be employed. However, even the latter
methods are not directly applicable to the problem studied here
because the redistribution integrals in the right-hand sides of the
four transfer equations~(\ref{PDE0})--(\ref{PDE3}) are not
forward-backward symmetric.  In this section I describe a variant of
the Feautrier method that has the desired properties for solving the
four one-dimensional radiative transfer
equations~(\ref{PDE0})--(\ref{PDE3}).  I follow in general the
procedure outlined by Milkey et al.\ (1975), pointing out the
differences that arise from the particular properties of the problem
studied here.

I choose as independent variables the Thomson scattering optical
depth
\begin{equation}
 \tau_{\rm es}(z)=\int_{0}^{z}n_{\rm e}(z)\sigma_{\rm T}dz\;,
 \label{taues}
\end{equation}
which is independent of photon energy, as well as the photon energy
$\epsilon$, and the direction of propagation $\mu$.  I then write the
four one-dimensional radiative transfer equations in the general form
\begin{equation}
\rho_{\rm s}\mu\frac{\partial I^m}{\partial \tau_{\rm es}}= 
 I^m-{\cal S}^{\rm m,s}-{\cal S}^{\rm m,a}\qquad m=0,..,3\;,
\label{igen}
\end{equation}
where ${\cal S}^{\rm m,s}$ and ${\cal S}^{\rm m,a}$ are the symmetric
and antisymmetric parts of the source functions and scattering integrals
in equations~(\ref{PDE0})--(\ref{PDE3}), which depend implicitly on
the radiation field.  Defining the Feautrier variables
\begin{equation}
u^m(\tau_{\rm es},\epsilon,\mu) \equiv \frac{1}{2}
  [I^m(\tau_{\rm es},\epsilon,\mu)+I^m(\tau_{\rm es},\epsilon,-\mu)]
\end{equation}
and
\begin{equation}
v^m(\tau_{\rm es},\epsilon,\mu) \equiv \frac{1}{2}
  [I^m(\tau_{\rm es},\epsilon,\mu)-I^m(\tau_{\rm es},\epsilon,-\mu)]\;,
\end{equation}
the transfer equation takes the form of the system of equations
(Milkey et al.\ 1975)
\begin{eqnarray}
\rho_{\rm s}\mu\frac{\partial u^{m}}{\partial \tau_{\rm es}}
   &=& 
 v^m-{\cal S}^{\rm m,s}
\label{ueq}\\
\rho_{\rm s}\mu\frac{\partial v^{m}}{\partial \tau_{\rm es}}&=& 
 u^m-{\cal S}^{\rm m,a}\;.
\label{veq}
\end{eqnarray}
Note, that because of the lack of forward-backward symmetry in the
scattering kernels, equations~(\ref{ueq})-(\ref{veq}) cannot be
combined into a single second-order equation, as in the usual
Feautrier method (see, e.g., Mihalas 1978).

I then discretize all quantities over $N_{\rm d}$ grid points in the
variable $\tau_{\rm es}$, $N_{\epsilon}$ grid points in photon energy,
and $N_\mu$ grid points in the direction of propagation $\mu$.  For
simplicity, I use, e.g., $u^m_{ij}$ to denote the first Feautrier
variable of order $m$ of the diffuse radiation field, evaluated on the
$i-$th grid point in the quantity $\tau_{\rm es}$, on the $k-$th grid
point in photon energy, and on the $l-$th grid point in the direction
of propagation $\mu$, such that $j=k+(l-1)N_\mu$.  In order to recover
the diffusion of photons in energy space because of Compton scattering
by thermal electrons, I use a second-order differencing scheme in
photon energy.  For example, I denote the first and second derivatives
of the first Feautrier variable with respect to energy by
\begin{equation}
\left.\frac{\partial u^m_{ij}}{\partial \epsilon}\right|_j=
\sum_{k=-1}^1 X_{jk}u^m_{i,j+k} 
\end{equation}
and
\begin{equation}
\left.\frac{\partial^2 u^m_{ij}}{\partial \epsilon^2}\right|_j=
\sum_{k=-1}^1 Y_{jk}u^m_{i,j+k}\;,
\end{equation}
where I have defined the operators
\begin{eqnarray}
X_{j,-1}&=&-\frac{1}{2}\left(\frac{1}{\epsilon_j-\epsilon_{j-1}}\right)\\
X_{j,0}&=&\frac{1}{2}\left(\frac{1}{\epsilon_j-\epsilon_{j-1}}
   -\frac{1}{\epsilon_{j+1}-\epsilon_j}\right)\\
X_{j,+1}&=&\frac{1}{2}\left(\frac{1}{\epsilon_{j+1}-\epsilon_j}\right)
\end{eqnarray}
and
\begin{eqnarray}
Y_{j,-1} & = & \frac{2}{(\epsilon_{j+1}-\epsilon_{j-1})(\epsilon_j-
   \epsilon_{j-1})}\\
Y_{j,0} & = & -2\left(\frac{1}{\epsilon_{j+1}-\epsilon_{j}}-
   \frac{1}{\epsilon_j-\epsilon_{j-1}}\right)
   \frac{1}{\epsilon_{j+1}-\epsilon_{j-1}}\\
Y_{j,+1} & = & \frac{2}{(\epsilon_{j+1}-\epsilon_{j-1})(\epsilon_{j+1}-
   \epsilon_j)}\;.
\end{eqnarray}
In differencing with respect to the variable $\tau_{\rm es}$, I note
that the quantity $u^m_{ij}$ is density-like and I, therefore, use
a center-differencing scheme for all interior grid points, i.e.,
\begin{equation}
  \left.\frac{\partial u^m_{ij}}{\partial \tau_{\rm es}}\right|_{i+1/2}=
  \frac{u^m_{i+1,j}-u^m_{i,j}}{\tau_{\rm es}^{i+1}-\tau_{\rm es}^{i}}\;.
\end{equation}
On the other hand, the quantity $v^m_{ij}$ is flux-like and I therefore
use for all interior grid points
\begin{equation}
  \left.\frac{\partial v^m_{ij}}{\partial \tau_{\rm es}}\right|_{i}=
  2\frac{v^m_{i+1/2,j}-v^m_{i-1/2,j}}
  {\tau_{\rm es}^{i+1}-\tau_{\rm es}^{i-1}}\;.
\end{equation}

Using the above differencing schemes, the difference equations in all interior
grid points become
\begin{eqnarray}
2(\rho_{\rm s})_{ij}\mu_j
\frac{1}{\tau_{\rm es}^{i+1}-\tau_{\rm es}^{i-1}}
  \left(v^m_{i+1/2,j}-v^m_{i-1/2,j}\right)&=& 
 u^m_{ij}-\sum_k R^{m,s}_{ijk}u^m_{ik}-G^{m,s}_{ij}
\label{veq_int}\\
(\rho_{\rm s})_{i+1/2,j}\mu_j
\frac{1}{\tau_{\rm es}^{i+1}-\tau_{\rm es}^{i}}
  \left(u^m_{i+1,j}-u^m_{ij}\right)&=& 
 v^m_{i+1/2,j}-\sum_k R^{m,a}_{i+1/2,jk}v^m_{i+1/2,k}-G^{m,a}_{i+1/2,j}\;,
\label{ueq_int}
\end{eqnarray}
where I have written explicitly the dependence of the source functions 
on the radiation field as
\begin{eqnarray}
{\cal S}^{m,s}_{ij}&=&\sum_k R^{m,s}_{ijk}u^m_{ik}+G^{m,s}_{ij}\\
{\cal S}^{m,a}_{i+1/2,j}
  &=&\sum_k R^{m,a}_{i+1/2,jk}v^m_{i+1/2,k}+G^{m,a}_{i+1/2,j}\;.
\end{eqnarray}
Note here that the above source functions have a different structure than
those of Milkey et al.\ (1975), because of the presence of the antisymmetric
term $G^a_{i+1/2,j}$ that does not depend on the diffuse radiation field.

At the illuminated surface of the slab, which I denote by $i=1$, the
boundary condition~(\ref{tau_bound}) translates into
$u^m_{1j}=v^m_{1j}$ and applying equation~(\ref{veq_int}) on the first half
of the first grid cell, denoted by $(1,3/2)$, I obtain
\begin{equation}
v^m_{3/2,j}=u^m_{1j}+\frac{1}{2\mu_j (\rho_{\rm s})_{1j}}
  (\tau_{\rm es}^2-\tau_{\rm es}^1)
 \left(u^m_{1j}-\sum_k R^{m,s}_{1jk}u^m_{1k}-G^{m,s}_{1j}\right)\;. 
 \label{bound1}
\end{equation}
At a very large optical depth, which I denote by $i=N_{\rm d}$, I set the flux
equal to zero, i.e., $v^{\rm m}_{N_dj}=0$, and applying
equation~(\ref{veq_int}) on the last half of the last grid cell,
denoted by $(N_{\rm d}-1/2,N_{\rm d})$, I obtain
\begin{equation}
v^m_{N_{\rm d}-1/2,j}=\frac{\tau_{\rm es}^{N_{\rm d}-1}-\tau_{\rm 
     es}^{N_{\rm d}}}{2(\rho_{\rm s})_{N_{\rm d}-1/2,j}\mu_j}  
     \left(u^m_{N_{\rm d},j}-\sum_k R^{m,s}_{N_{\rm d}jk}u^m_{N_{\rm 
     d}k}-G^{m,s}_{N_{\rm d}j}\right)\;. 
  \label{bound2}
\end{equation}
Note that this boundary condition is different than
equation~(\ref{tau_bound}).  Finally, at the first and last energy
grid-points, boundary conditions~(\ref{en_bound}) become simply
\begin{equation}
u^m_{i,1+(l-1)N_\mu}=u^m_{i,N_\epsilon+(l-1)N_\mu}=
v^m_{i,1+(l-1)N_\mu}=v^m_{i,N_\epsilon+(l-1)N_\mu}=0\;.
 \label{bound3}
\end{equation}

The system of algebraic equations~(\ref{ueq_int})--(\ref{bound3}) 
can be written in the general matrix form 
\begin{eqnarray}
A^m_i u^m_i + B^m_i u^m_{i+1}&=&C^m_i v^m_{i+1/2}-G^{m,a}_i
    \qquad\;, i=1,\ldots,N_d-1\nonumber\\
D^m_i v^m_{i-1/2} + E^m_i v^m_{i+1/2}&=&F^m_i u^m_i-G^{m,s}_i
    \qquad\qquad\;, i=1, \ldots, N_d\;, 
 \label{system}
\end{eqnarray}
where $A^m_i$, $B^m_i$, $D^m_i$, and $E^m_i$ are $N_\epsilon 
N_\mu\times N_\epsilon N_\mu$ diagonal matrices, $C^m_i$ and $F^m_i$ 
are full $N_\epsilon N_\mu\times N_\epsilon N_\mu$ matrices, 
and $G^{m,a}_i$ and $G^{m,s}_i$ are $N_\epsilon N_\mu\times 1$ vectors.
Equations~(\ref{system}) can then be solved recursively from $i=N_{\rm 
d}$ to $i=1$, using the relations
\begin{eqnarray}
u^m_i &=& V^m_i v^m_{i+1/2}+W^m_i\\
v^m_{i+1/2}&=&U^m_{i+1/2} u^m_{i+1}+T^m_{i+1/2}\;,
\end{eqnarray}
where
\begin{eqnarray}
V^m_i & = & (F^m_i - D^m_i U^m_{i-1/2})^{-1}E^m_{i}\\
W^m_i & = & (F^m_i - D^m_i U^m_{i-1/2})^{-1}(D^m_iT^m_{i-1/2}+G^{m,s}_i)\\
U^m_{i+1/2} & = & (C^m_i - A^m_i V^m_i)^{-1}B^m_{i}\\
T^m_{i+1/2} & = & (C^m_i - A^m_i 
   V^m_i)^{-1}(A^m_iW^m_{i-1/2}+G^{m,a}_{i-1/2})\;.
\end{eqnarray}
Note here a misprint in the elimination scheme of Milkey et al.\ (1975) 
as well as a difference with the above scheme that arises because
of the presence of the antisymmetric term $G^{m,a}_{i-1/2}$.

It is important to point out here that the method presented in this
section is very efficient since it requires the solution of only four
equations of simplicity equal to the minimum required for calculating
the interaction of an external radiation field with an accretion disk.
Therefore, the computational cost of this method is exactly equal to
four times the cost required for solving the simplest problem of
normal illumination of a plane-parallel slab. As an example, the
solution of a problem with 20 grid points in optical depth, 10 grid
points in angle, and 60 grid points in energy (i.e., similar to the
one shown in Fig.~1) requires only 2~CPU minutes on a 500~MHz Alpha
processor.

\begin{figure}[t]
  \centerline{
   \psfig{file=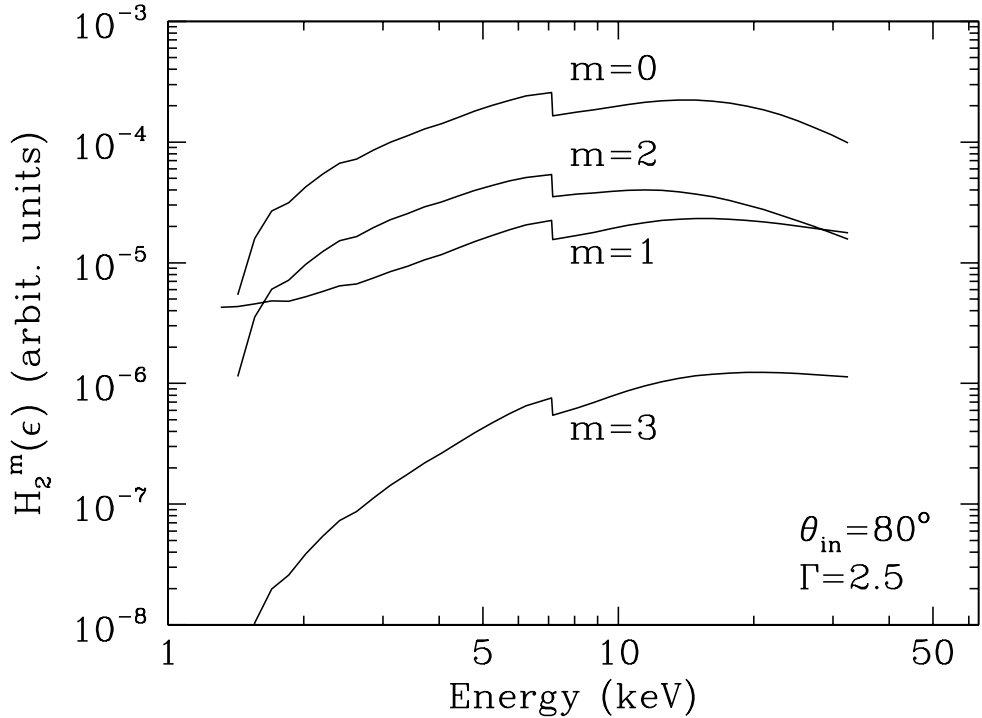,angle=-90,width=14truecm}} 
\caption{\footnotesize 
The energy dependence of the vertical flux emerging from an
illuminated accretion disk, decomposed into the various expansion
orders of the specific intensity. The angle of illumination is set to
$\theta_{\rm in}=80^\circ$ with respect to the normal and the photon
index of the illuminating spectrum to $\Gamma=2.5$.}
\label{fig:energy}
\end{figure} 

\section{APPLICATION TO ILLUMINATED, COLD ACCRETION DISKS}

In this section, I use the method described in \S3 for calculating the
albedo of an illuminated, cold accretion disk.  I consider concentric
annuli of the disk, which I approximate by plane-parallel slabs.  I
take into account Compton scattering as well as bound-free absorption
from a cold gas (Morrison \& McCammon 1983) and neglect all other
radiation processes.  When the radiation field is weak, the ratio of
the bound-free absorption coefficient to the scattering cross section
is independent of the electron density and temperature.  As a result,
I can solve the radiative transfer problem in terms of the
electron-scattering optical depth $\tau_{\rm es}$ without the need to
consider the vertical disk structure.  When the radiation field is
strong, the illumination of the disk will affect both its ionization
balance and the absorption coefficients, and hence the overall
solution will depend explicitly on the vertical structure of the disk
(see, e.g., Nayakshin \& Kallman 2001).

\begin{figure}[t]
  \centerline{
   \psfig{file=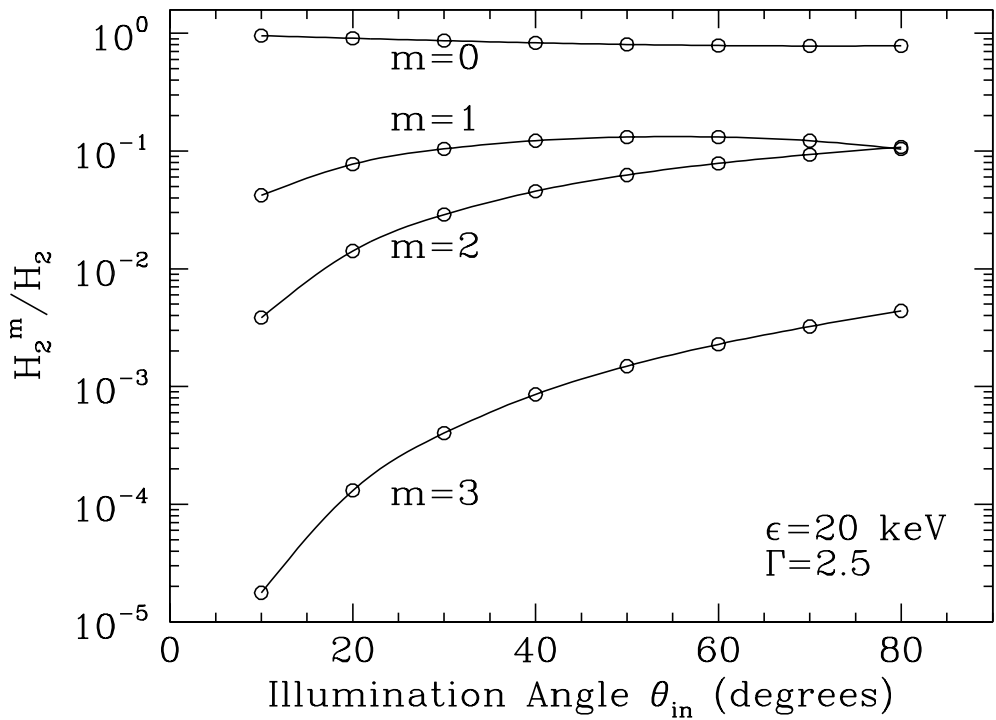,angle=-90,width=14truecm}} 
\caption{Relative contribution to the vertical flux of the various
expansion orders of the specific intensity, for different angles of
illumination. The vertical flux is calculated for a photon energy of
20~keV and a power-law spectrum of illumination with a photon index
$\Gamma=2.5$.}  
\label{fig:contrib}
\end{figure} 

In general, the radiative transfer equation~(\ref{4dRTE}) is linear in
the specific intensity. I can, therefore, write
 \begin{equation}
 I(\tau,\mu,\phi,\e)=I_{\rm v}(\tau,\mu,\phi,\e)
   +I_{\rm i}(\tau,\mu,\phi,\e)\;,
 \label{dec}
 \end{equation}
where $I_{\rm v}$ and $I_{\rm i}$ describe the radiation field due to
viscous heating and illumination respectively, and solve for the two
radiation fields independently. Note here that
decomposition~(\ref{dec}) is only formally valid, since the absorption
and emission coefficients, as well as the scattering kernel may depend on
the electron density and temperature that are determined by the total
radiation field. The model parameters for the calculations include the
angle of illumination, $\theta_{\rm i}$ with respect to the normal and
the spectrum of the illuminating radiation, which I set to a power-law of
photon index $\Gamma$, i.e., $F(\epsilon)\sim \epsilon^{-\Gamma+1}$.

Figure~1 shows the contribution to the vertical flux (Eq.~[\ref{H2}])
of the various orders of the decomposition~(\ref{expans}), for an
illumination angle of $\theta_{\rm i}=80^\circ$ and a power-law
spectrum of photon index $\Gamma=2.5$. Even though the term of
zeroth-order provides the dominant contribution to the vertical flux,
the contribution of higher-order terms is not negligible. This is
shown in Figure~2, where the contribution of the high-order terms to
the vertical flux is plotted for different angles of illumination. The
total correction, caused by the obliqueness of the illumination,
ranges between $\sim 5-20$\%. Note here that the relative
contributions of the different orders of the specific intensity
plotted in Figure~2 correspond to the case of zero electron
temperature and increase for finite electron temperatures because of
the asymmetry of the Compton terms in the scattering kernel.

In the context of accretion onto galactic compact objects, the albedo
$\eta$ of an accretion disk is usually defined in terms of the
fraction of the illuminating flux that does not heat the disk gas
(see, e.g., de Jonk et al.\ 1996). In this section, I first give the
general expression for the albedo of an accretion disk that is
illuminated obliquely and then evaluate it for the case of
geometrically thin, optically thick, cold accretion disks.

Starting from equation~(\ref{dec}) and integrating the transfer equation 
for $I_{\rm i}$ over photon energy and directional angle I obtain
 \begin{equation}
 \nabla\cdot \vec{H}_{\rm i}=
 -\chi J_{\rm i}-n_e\sigma_{\rm T}
  \left(4\frac{T_{\rm e}}{m_{\rm e}}-
       \frac{\langle\e\rangle}{m_{\rm e}}\right)
    J_{\rm i}\;,
 \end{equation}
 where $J_{\rm i}$ and $\vec{H}_{\rm i}$ are the
energy-integrated, zeroth and first angular moments of
$I_{\rm i}$ and 
 \begin{equation}
 \langle\e\rangle\equiv \frac{
 \int\e J_{\rm i}(\e) d\e }{
 \int J_{\rm i}(\e) d\e}\;.
 \end{equation}
 The rate $Q_{\rm i}$, at which the illuminating radiation heats
the gas is therefore
 \begin{equation}
 Q=4\pi\left[\chi +n_e\sigma_{\rm T}
  \left(
       \frac{\langle\e\rangle}{m_{\rm e}}
      -4\frac{T_{\rm e}}{m_{\rm e}}\right)
    \right]J_{\rm i}\;,
 \label{Q}
 \end{equation}
and the albedo can be written in terms of the volume integral of $Q$ as
 \begin{eqnarray}
 \eta&=&1-\frac{4}{\mu_{\rm i}}\int_0^{\tau_{\rm es,max}}
  \left[\chi +n_e\sigma_{\rm T}
  \left(
       \frac{\langle\e\rangle}{m_{\rm e}}
      -4\frac{T_{\rm e}}{m_{\rm e}}\right)
    \right]\frac{J_{\rm i}}{F} d\tau_{\rm es}\nonumber\\
   &=&1-\frac{2}{\mu_{\rm i}F}\int_0^{\tau_{\rm es,max}}d\tau_{\rm es}
  \left[\chi +n_e\sigma_{\rm T}
  \left(
       \frac{\langle\e\rangle}{m_{\rm e}}
      -4\frac{T_{\rm e}}{m_{\rm e}}\right)
    \right]\int_\epsilon d\epsilon \int_{-1}^1 d\mu
   \I^0(\tau_{\rm es},\e,\mu)\;,
  \label{albedo}
 \end{eqnarray}
 where $F$ is the energy-integrated flux of the illuminating radiation and
$\tau_{\rm es,max}$ is the vertical height of the accretion disk. 
Expression~(\ref{albedo}) allows the study of the vertical profile of the
heating of the accretion disk by the irradiating atmosphere as well as 
the calculation of its albedo, by solving {\em only\/} the zeroth-order
transfer equation for $\I^0(\tau_{\rm es},\e,\mu)$.

\begin{figure}[t]
  \centerline{
   \psfig{file=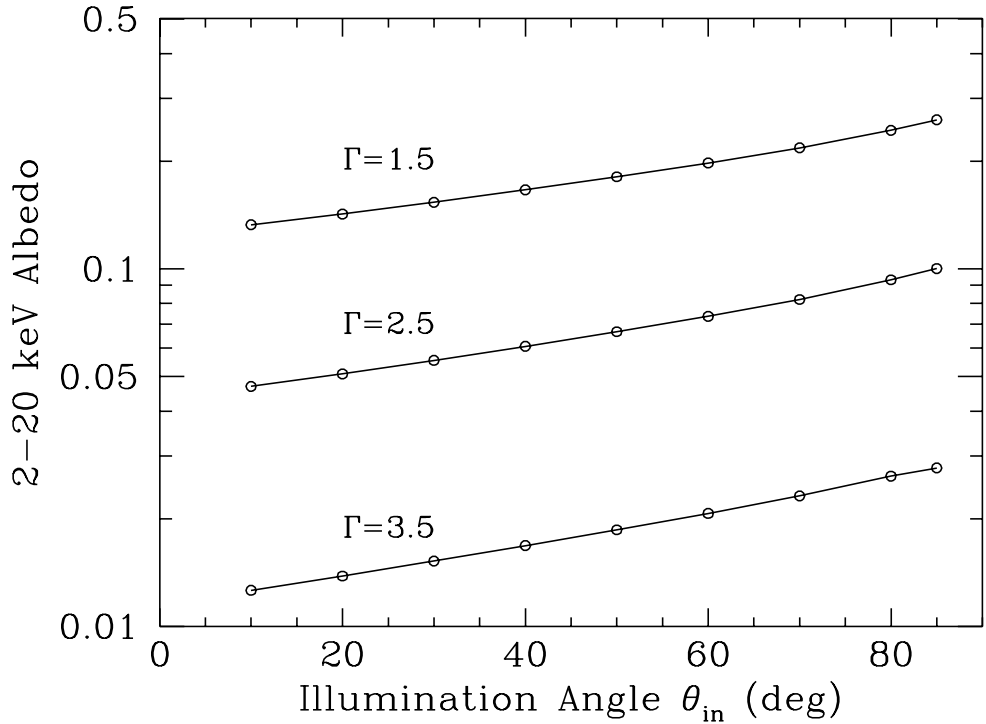,angle=-90,width=14truecm}} 
\caption{The albedo of a cold accretion disk, calculated in the
  $2-20$~keV energy range, for different power-law spectra and angles of
  illumination.}  
\label{fig:albedo}
\end{figure} 

Figure~3 shows the albedo of a cold accretion disk, calculated in the
$2-20$~keV energy range, for different power-law spectra and angles of
illumination.  As expected, for larger photon indices of the
illuminating radiation, the fraction of low-energy photons, which are
efficiently absorbed, is higher and hence the albedo of the disk is
lower. At the same time, as the angle of illumination increases, the
photons interact with the electrons in a shallower layer of the
accretion disk and therefore have a higher chance of escaping after
one interaction, increasing the value of the albedo. The overall
effect of our treatment of the obliqueness of irradiation is this
systematic increase of the disk albedo with illumination angle, which
can be up to a factor of $\sim 2$ larger that in the case of normal
illumination.

\section{CONCLUSIONS}

In this paper, I studied the transfer of radiation in an accretion
disk that is obliquely illuminated by an external source of
radiation. I showed that the resulting transport problem can be
decomposed exactly into four one-dimensional problems, which I solved
using a variant of the Feautrier method. I then applied this method in
calculating the albedos of cold accretion disks.

The calculated values for the albedos are $\le 0.30$, even for
the softer spectra and larger illumination angles considered
here. These values are small and cannot account, for example, for the
observed optical magnitudes of galactic low-mass X-ray binaries, which
require albedos in excess of $\simeq 0.90$ (see de Jong et al.\
1996). Figure~3 shows that relying on a near-grazing illumination of
the accretion disk is not enough to account for the observations. High
ionization fractions at the surface layers of the disk, which would
reduce the absorption of photons, or even the existence of a highly
ionized scattering wind above the accretion disk is probably required
for the calculated albedos to reach the high values inferred from
observations.

In this study, I assumed for simplicity that all heavy elements in the
accretion disk are neutral and, therefore their interaction with the
illuminating photons is described by the bound-free opacities of
Morrison \& McCammon (1983). In reality, however, the heated skin of
the accretion disk will be collisionally- and photo-ionized and its
vertical ionization and thermal balance will need to be calculated
self-consistently with the radiation field (e.g., Raymond 1993; Ko \&
Kallman 1994). Note, however, that the calculation of both the
ionization balance and the radiative equilibrium depend only on the
zeroth moment of the specific intensity (see eq.~[\ref{J}] and
[\ref{Q}]) and, therefore, require the solution of only the
zeroth-order transfer equation. As a result, the properties of the
disk gas can be calculated exactly in a simple, one-dimensional
configuration and the full angular dependence of the radiation field
can then be calculated with prescribed gas properties. The results of
a self-consistent caclulation of the radiation and gas properties will
be reported elsewhere.

\acknowledgements

I am grateful to G.\ Rybicki for bringing to my attention the
possibility of decomposing a multi-dimensional transfer equation into
a small number of one-dimensional equations and for carefully reading
the manuscript.  I also thank Feryal \"Ozel for many useful
discussions, especially on the implementation of the Feautrier method
in problems with no forward-backward symmetry. This work was supported
by a postdoctoral fellowship of the Smithsonian Institution and also,
in part, by NASA.

\end{document}